\documentclass{emulateapj}
\usepackage{graphicx}
\usepackage{t1enc}
\usepackage{apjfonts}

\newcommand{\kms}   {km~s$^{-1}$}
\newcommand{\cmt}   {cm$^{-3}$}
\newcommand{\lo}    {$L_{\sun}$}
\newcommand{\mo}    {$M_{\sun}$}
\newcommand{\nh}    {NH$_3$}
\newcommand{\et}    {et al.}
\newcommand{\eg}    {e.\,g.,}
\newcommand{\jpb}   {$\rm Jy~beam^{-1}$}	

\begin{document}

\title{
Submillimeter emission from the hot molecular jet HH\,211
}

\author{
A. Palau\altaffilmark{1}, 
P. T. P. Ho\altaffilmark{2}, 
Q. Zhang\altaffilmark{2}, 
R. Estalella\altaffilmark{1}, 
N. Hirano\altaffilmark{3},
H. Shang\altaffilmark{3}, 
C.-F. Lee\altaffilmark{2}, \\
T. L. Bourke\altaffilmark{2},  
H. Beuther\altaffilmark{2,4}, \&
Y.-J. Kuan\altaffilmark{5,3} 
}

\altaffiltext{1}{ 
Departament d'Astronomia i Meteorologia, Universitat de Barcelona,
Av.\ Diagonal 647, E-08028 Barcelona, Spain}

\altaffiltext{2}{ 
Harvard-Smithsonian Center for Astrophysics,
60 Garden Street, Cambridge, MA 02138, USA}

\altaffiltext{3}{
Academia Sinica, Institute of Astronomy \& Astrophysics, P.O. Box
23--141, Taipei, 106, Taiwan, R.O.C.} 

\altaffiltext{4}{
Max-Planck-Institut for Astronomy, Koenigstuhl 17, 69117 Heidelberg, Germany
}

\altaffiltext{5}{
Departament of Earth Sciences, National Taiwan Normal University, 88
Sec. 4, Ting-Chou Rd., Taipei, 116, Taiwan, R.O.C.}

\shortauthors{Palau \et}

\shorttitle{Submillimeter Emission from HH\,211}

\begin{abstract}

We observed the HH\,211 jet in the submillimeter continuum and the CO(3--2) and
SiO(8--7) transitions with the Submillimeter Array.  The continuum source
detected at the center of the outflow shows an elongated morphology,
perpendicular to the direction of the outflow axis. The high-velocity emission
of both molecules shows a knotty and highly collimated structure.  The
SiO(8--7) emission at the base of the outflow, close to the driving source,
spans a wide range of velocities, from $-20$ up to 40 \kms. This suggests that
a wide-angle wind may be the driving mechanism of the HH\,211 outflow.  For
distances $\ge 5''$ ($\sim 1500$ AU) from the driving source, emission from
both transitions  follows a Hubble-law behavior, with SiO(8--7) reaching higher
velocities than CO(3--2), and being located upstream of the CO(3--2) knots.
This indicates that the SiO(8--7) emission is likely tracing entrained gas very
close to the primary jet, while the CO(3--2) is tracing less dense entrained
gas.  From the SiO(5--4) data of Hirano et al.\ we find that the
SiO(8--7)/SiO(5--4) brightness temperature ratio along the jet decreases for
knots far from the driving  source. This is consistent with the density
decreasing along the jet, from  (3--10)$\times 10^6$ \cmt\ at  500 AU to
(0.8--4)$\times 10^6$ \cmt\ at 5000 AU from the driving source.

\end{abstract}


\maketitle

\section{Introduction}

HH\,211 is a warm and energetic molecular outflow located in the IC\,348
complex at 315 pc, which was discovered by McCaughrean \et \ (1994) from
observations of H$_2$ (at 2.12 $\mu$m). The inclination from the plane of the
sky is supposed to be small, around 10$\degr$ (Hirano \et\ 2005).  The outflow
in the CO(2--1) transition shows a well collimated structure at high 
velocities, and traces the outflow cavity walls at low velocities (Gueth \& 
Guilloteau 1999, hereafter GG99). On the other hand, the SiO emission has been
detected  with single-dish telescopes up to the 11--10 transition, indicative
of gas densities $> 10^6$ \cmt\ along the jet (Nisini \et \ 2002; Chandler \&
Richer 2001). Recently, Hirano \et \ (2005) observed HH\,211 in the SiO(5--4)
transition and found a highly collimated structure consisting of a chain of
knots. The innermost knots likely trace the primary jet launched at the close
vicinity of the protostar. In order to follow up the study of the excitation
conditions along the outflow, we have carried out observations of the high-J
transitions CO(3--2) at 345.796 GHz and SiO(8--7) at 347.331 with high-angular
resolution.

\section{Observations} 
 
Observations of HH\,211 with the Submillimeter Array\footnote{The Submillimeter
Array is a joint project between the Smithsonian Astrophysical Observatory and
the Academia Sinica Institute of Astronomy and Astrophysics, and is funded by
the Smithsonian Institution and the Academia Sinica.} (SMA; Ho \et \ 2004) in
the 345 GHz band  were carried out on 2004 October 4 and October 18, with
seven  antennas for each day in the compact configuration. The phase reference
center  of the observations was $\alpha(\mathrm J2000)=03^{\mathrm
h}43^{\mathrm m}56\fs8$, $\delta(\mathrm J2000)=+32\degr00\arcmin50\farcs4$,
and the projected baselines ranged from 14.7 to 127 m. The half-power width of
the SMA primary beam at 345 GHz is $\sim 36''$. During each track, we observed 
3 pointing fields, separated from the central pointing by $26''$ (to southeast)
and $21''$ (to northwest) along the axis of the jet. The correlator  was
configurated in the standard mode, providing a uniform spectral resolution 
across the full 2 GHz IF band in each sideband of 0.8125 MHz (or 0.7 \kms). The
frequency range covered by the lower and upper sidebands was 335.58--337.55
GHz, and 345.59--347.56 GHz, respectively. The passband of  each dataset was
calibrated in MIR-IDL by using  both sidebands of Saturn for  the Oct 4
dataset, and the lower sideband of Venus for the Oct 18 dataset (lower sideband
of Venus was used to calibrate the lower sideband and the upper sideband).  The
maximum error due to passband calibration across the full 2 GHz sideband was
about $\sim 20$\%. Gain calibration of the visibility phases and amplitudes and
flux calibration was  performed in MIRIAD  using 3C 84 as the gain and flux
calibrator, for which we set the flux to be 1.7 Jy  (value independently
measured with the SMA at 345 GHz within 15 days from our  observations).  The
typical rms of the gain phases was $\sim 65 \arcdeg$, and  the overall flux
uncertainty is estimated to be about 15\%. 

Imaging of data cubes was made in MIRIAD by combining the data from the 3 
pointings in the visibility plane, and cleaning in a box covering the full
extent of  the jet. The final maps include both Oct 4 and Oct 18 datasets, and
have a synthesized beam of $1\farcs94 \times 0\farcs97$ (PA$=67\fdg2$), with 
an rms noise level per channel (of 2 \kms \ wide) of 0.30 and 0.25 \jpb \ for
the CO(3--2) and SiO(8--7), respectively.  Although weather conditions were not
very good on Oct 4 ($\tau_{230} \simeq 0.1$, compared with $\tau_{230} \simeq
0.04$ for Oct 18), the combination of Oct 4 and Oct 18 datasets improved the
S/N for the line emission. Continuum emission at 345 GHz was obtained by
averaging the  spectral channels free of line emission.  The continuum image
shown in this letter is the result of combining the upper and lower sideband
data of the Oct 18 dataset only, with natural weighting of the uv data and
without spatial restrictions in the image cleaning process. The synthesized
beam of the continuum map was $2\farcs04 \times 1\farcs02$ (PA$=63\fdg9$), and 
the rms noise level was 7 m\jpb.

\section{Results} 
 
\subsection{CO(3--2) and SiO(8--7) emission \label{srcosio}} 
 
The systemic velocity of HH\,211 is 9.2 \kms \  (velocities in this paper are
LSR). CO(3--2) emission is detected up to $-14$ \kms \ for the blue side of
the  outflow, and up to 40 \kms \ for the red side.  In Fig.\ \ref{fm0}a we
show an overlay of the low-velocity CO(3--2)  emission  on the H$_2$ emission
at 2.12 $\mu$m from Hirano \et \ (2005). We find CO(3--2) low-velocity emission
associated  with the brightest infrared knots, as well as weak emission tracing
a shell-like structure, similar to the CO(2--1) emission in the same velocity
range (GG99). 

The CO(3--2) high-velocity emission ("high-velocity" refers to velocitites
lower than 0 \kms\ and higher than 20 \kms) traces a very well collimated and
knotty jet-like structure, and  is very close to the axis of the cavity traced
by the  low-velocity CO and the H$_2$ emission (see Fig.\ \ref{fm0}b).

The SiO(8--7) emission is very weak for velocities close to the systemic 
velocity, but is detectable up to $-20$ and 42 \kms\  (see spectrum in Fig.
\ref{fm0}c).    The high-velocity SiO(8-7) emission (Fig.\ \ref{fm0}c) is also 
very well collimated and clumpy. However, the SiO is  found much closer to the
protostar than the CO for the  high-velocity emission.   We have adopted the
same nomenclature for the knots as Hirano  \et \  (2005). Toward the strongest 
H$_2$ features, there is no significant  high-velocity CO(3--2) or SiO(8--7)
emission (for a discussion on H$_2$ excitation and a comparison with SiO
emission in HH\,211, see Chandler \& Richer 2001). Regarding the medium in which
the HH\,211 jet is propagating, the emission from \nh \ and H$^{13}$CO$^+$
reveals an elongated condensation of $\sim 1$ \mo, located on the red side of
the jet, as shown in Fig.\ \ref{fm0}b (Bachiller \et\ 1987; GG99; Wiseman \et \
2001). 

We estimated the physical parameters of the outflow from the CO(3--2) emission
(Table~\ref{toutpar}), following Yang \et\  (1990) and assuming optically thin
emission in the wing, and an excitation temperature of $\sim 12$ K (derived
from the line intensity). The values shown in the Table are not corrected for
the inclination effect. The derived age and mass are consistent with those
obtained by GG99 from CO(1--0) and CO(2--1) with the PdBI.

\begin{deluxetable*}{ccccccccc} 
\tablecaption{Physical Parameters of the HH\,211 Outflow from CO(3--2) \label{toutpar}} 
\tablehead{ 
\colhead{Age}  
&\colhead{$N_{12}$} 
&\colhead{Mass} 
&\colhead{$\dot{M}$} 
&\colhead{$P$} 
&\colhead{$\dot{P}$}
&\colhead{$E_{\mathrm{kin}}$} 
&\colhead{$L_{\mathrm{mech}}$}
&\colhead{$L_\mathrm{bol}$}\\ 
\colhead{(yr)} 
&\colhead{(cm$^{-2}$)} 
&\colhead{(\mo)} 
&\colhead{(\mo~yr$^{-1}$)} 
&\colhead{(\mo~\kms)} 
&\colhead{(\mo~\kms~yr$^{-1}$)}
&\colhead{(erg)} 
&\colhead{(\lo)} 
&\colhead{(\lo)}
} 
\startdata 
1400 
&1.4$\times10^{17}$ 
&0.0024 
&1.7$\times10^{-6}$
&0.040
&2.8$\times10^{-5}$
&6.9$\times10^{42}$
&0.027
&3.6\tablenotemark{a}
\enddata 
\tablenotetext{a}{From Froebrich (2005).}
\end{deluxetable*}

\subsection{Continuum emission \label{srcont}} 
 
In Fig.\ \ref{fm0}b, we show the continuum emission overlaid on the 
high-velocity CO(3--2) emission. We detected the source at the center of the 
outflow with $S/N=18$. The position for the source derived from a Gaussian fit 
is  $\alpha(\mathrm J2000)=03^{\mathrm h}43^{\mathrm m}56\fs8$, $\delta(\mathrm
J2000)=+32\degr00\arcmin50\farcs2$. The deconvolved size of the source is
$(1\farcs6 \pm 0\farcs2) \times (0\farcs6 \pm 0\farcs1)$,  corresponding to
$510 \times 200$ AU, and the deconvolved position angle (PA)  is $26\degr \pm
4\degr$.  The PA of the large-scale jet in the integrated SiO  emission  has
been determined to be $116\fdg2 \pm 0\fdg2$. Thus, we  find that the
submillimeter continuum source, which is likely tracing the disk, is  exactly
perpendicular to the large-scale collimated jet emanating from it, to within
the measurement error. 

The peak intensity of the source is $0.13 \pm 0.02$ \jpb, and the flux density
is $0.22 \pm 0.02$ Jy.  Assuming that the dust emission is optically thin and
well described by a modified black body law, we  can estimate the mass of the
disk, for a given dust emissivity index and dust  temperature. We assumed the
opacity law of Beckwith \et \ (1990), and an emissivity index $\beta \simeq 1$.
Then, for dust temperatures of 20--40 K, we derive a mass for the disk ranging
from 0.02 to 0.06 \mo, only $\sim 5$\% of the mass of the large-scale \nh\ 
condensation (\S\ \ref{srcosio}). The derived mass is 2--3 times lower than the
values obtained at 230 GHz by GG99 and Hirano \et\ (2005), indicating that our
measurements at 345 GHz could be tracing only the warmer part of the dusty
disk.

\section{Discussion and Conclusions} 

\subsection{SiO(8--7) versus SiO(5--4) \label{sdsiosio}} 
 
Observations toward HH\,211 in the SiO(5--4) transition at 217 GHz were carried 
out by Hirano \et \  (2005), with similar angular resolution. In order to
compare the maps from both transitions, we convolved the moment-zero maps
(integrated over all velocities) with a Gaussian to achieve a final beam of 
$1\farcs95$, the largest major axis of the SiO(8--7) and SiO(5--4) beams.

\begin{figure}[ht] 
\epsscale{0.8}
\plotone{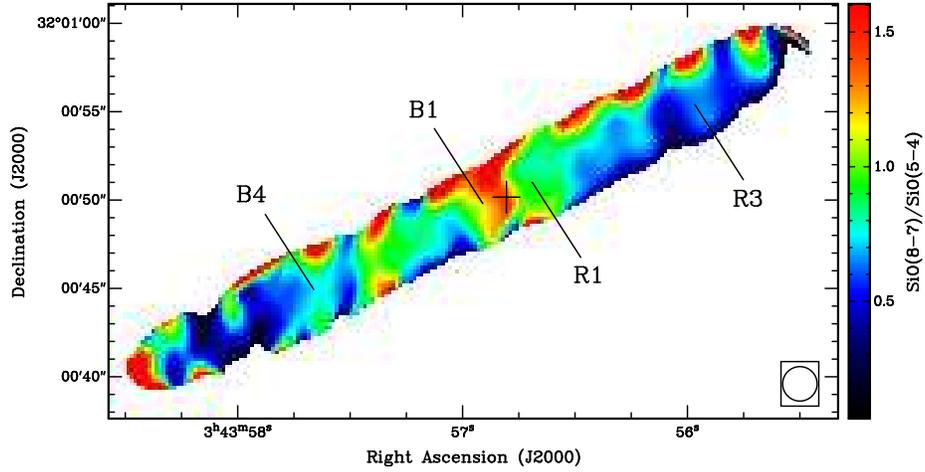} 
\caption{ 
Map of the ratio SiO(8--7)/SiO(5--4), performed using the task MATHS in MIRIAD 
on the images of SiO(8--7) and SiO(5--4) integrated over all the velocity 
range, and smoothed to the same angular resolution of $1\farcs95$. The ratio
is made by shifting the SiO(5--4) map ($+0\farcs19,  +0\farcs07$), the
offset between the positions of the continuum source at 217 GHz  and 345 GHz.
\label{sio8754ratiomap} 
} 
\end{figure}

We computed the SiO(8--7)/SiO(5--4) ratio map after correcting for the offset
found  in the position of the continuum source of both images ($\sim
0.2''$ in declination), and the  result is shown in Fig.\ \ref{sio8754ratiomap}.
The uncertainty in the ratio is $\sim 20$\%. While the value for the ratio at
the position of the innermost knots, B1 and R1, is $\sim 1$, far away from the
protostar the  ratio decreases down to $\sim 0.5$. Comparing the ratio with the
results of LVG modeling of Nisini \et\  (2002),  we set ranges for the density.
Note that the  SiO(8--7)/SiO(5--4) ratio is not very sensitive to temperature
variations for  $T \gg 100$ K, since  the SiO(8--7) upper level energy is $\sim
75$ K. At the position of B1 and R1 ($\sim 500$ AU) we estimate that, for
temperatures in the range 100--1000 K, the density must be (3--10)$\times 10^6$
\cmt.  On the other hand, $\sim 15''$ (or 5000 AU) away from the center of the
jet the ratio is 0.5, and this yields a density of (0.8--4)$\times 10^6$ \cmt.
This is consistent with the density derived by Hirano \et \ (2005) from the
SiO(5--4) jet and by Nisini \et \  (2002) from the single-dish data. Therefore,
the density of the innermost knots seems to be about one order of magnitude
higher than that of the knots further out along the jet.

\subsection{SiO(8--7) versus CO(3--2)  \label{sdsioco}} 

\begin{figure}[ht] 
\epsscale{0.7}
\plotone{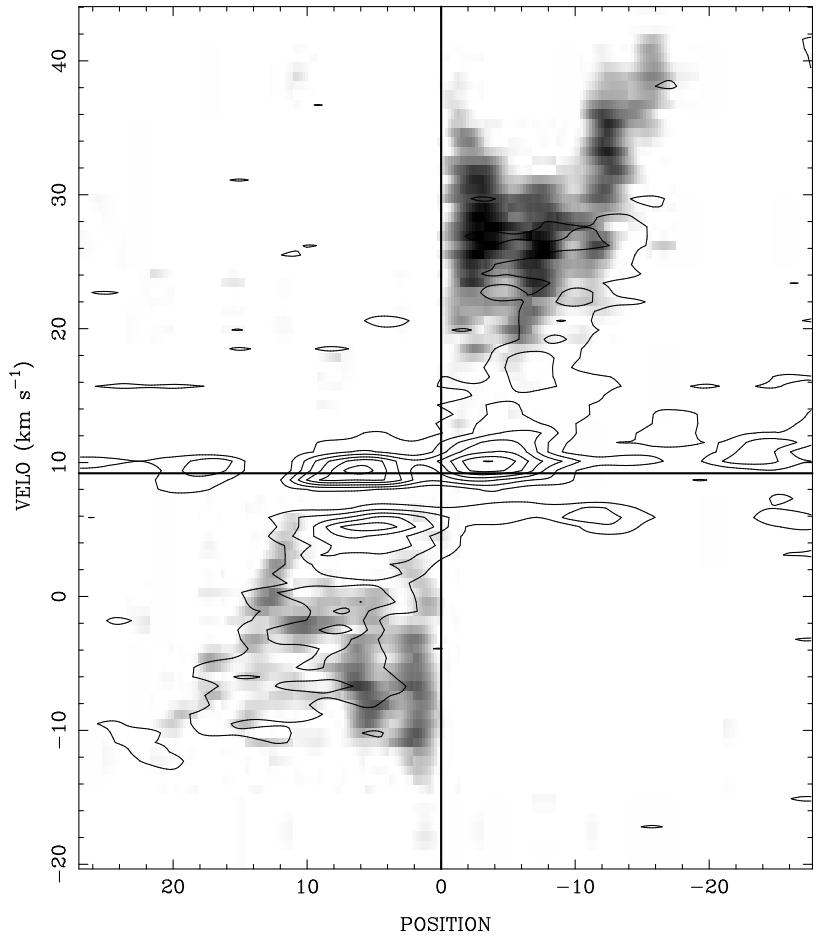} 
\caption{ 
CO(3--2) position-velocity (p-v) plot (contour) overlaid on the SiO(8--7) (grey
scale).   Velocity resolution is  0.7 \kms. Position is in units of arcsec. The
CO(3--2) plot was computed by smoothing the channel maps  with a Gaussian of 
$4'' \times 2''$ and PA$=26 \arcdeg$. The range of the  grey scale is from 0.6
to 2.8 \jpb, and contours start at 0.55 \jpb \ and  increase in steps of 0.55
\jpb. The straight lines indicate the  position of the continuum source and the
systemic velocity. \label{fpv}
} 
\end{figure}

Position-velocity (p-v) plots were computed for both CO(3--2) and  SiO(8--7)
emission from 0.7 \kms \ wide channel maps along the axis of the jet,  PA$=116
\arcdeg$. Since the CO emission is somewhat extended in the  direction
perpendicular to the jet axis, we have smoothed the image with a  gaussian twice
the beam of the observations, and with a position angle  perpendicular to the
axis of the jet, in order to enhance the S/N of the CO emission in the p-v
plot.  An overlay of SiO on the CO p-v plot is presented in  Fig.\ \ref{fpv}. 

First of all, there is a distinct gap in CO emission from 7 to 8  \kms,
affecting all positions  along the jet. CO(3--2) and (1--0) observations of low
angular resolution ($\sim 15''$; Hirano \et, in preparation) show an absorption
feature at the same velocity. The gap is probably  due to an  intervening cold
cloud along the line of sight. 

\begin{figure*}[ht]
\epsscale{0.8}
\plotone{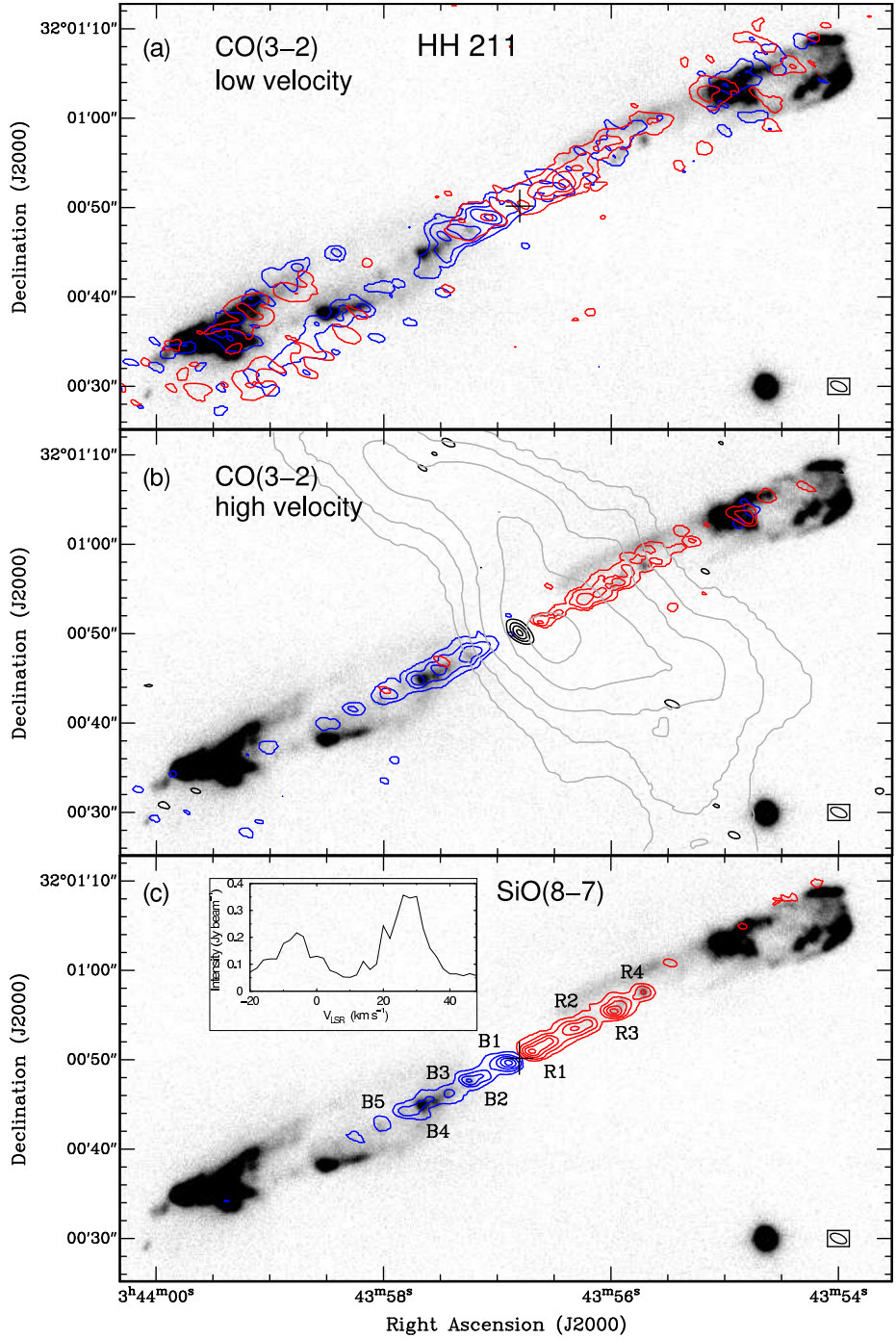}
\caption{\small
\emph{(a)} Contours: CO(3--2) emission  integrated for low velocities (2 to 8
\kms, blue, and 10 to 18 \kms, red). Blue and red contours start at 15\% of the
peak  intensity (blue: 26.34 \jpb \ \kms; red: 33.44 \jpb \ \kms), and increase
in 25\% intervals. The cross  marks the position of the submillimeter continuum
source (as in the bottom panel).  
\emph{(b)} Color contours: CO(3--2) emission  integrated from $-14$ to 0 \kms \
(blue) and from 20 to 40 \kms \ (red). Blue and red contours start at 10\% of
the peak intensity (blue: 25.31 \jpb \ \kms; red: 18.36 \jpb \ \kms), and
increase in steps of 30\%.  Black contours: submillimeter continuum emission.
Contours start at 0.03 \jpb  \, and increase in steps of 0.03 \jpb. Grey
contours: zero-order moment map of the  NH$_3$(1,1) emission from Wiseman \et \
(2001). Contours start at 10\% of the  peak intensity, 94.9 \jpb \  \kms, and
increase in steps of 20\%.
\emph{(c)} Contours: SiO(8--7) emission integrated from $-20$ to 0 \kms \
(blue) and from 20 to 42 \kms \ (red). Blue and red contours start at 10\% of
the peak intensity (blue: 41.75 \jpb \ \kms; red: 49.09 \jpb \ \kms), and
increase in 20\% intervals. Knots labeling is the same as Hirano \et \ (2005).
An inlay of the SiO(8--7) spectrum averaged over the central $10''$ of the jet
is also shown (1 \jpb\ corresponds to 5.4 K). In all panels, the grey scale is
the deep image of the H$_2$ emission at $2.12 ~\mu$m from Hirano \et \ (2005),
and the clean beam, shown in the bottom-right corner, is $1\farcs94 \times
0\farcs97$, PA$=67\fdg2$. \label{fm0} 
}   
\end{figure*}

In the CO(3--2) p-v plot, one can see a low-velocity component, extending along
all positions and  tracing the shell structure seen in the low-velocity map from
Fig.\ \ref{fm0}a,  and a second component tracing the high-velocity material,
with velocities  increasing with distance from the protostar (Hubble-law), up to
velocities of  $\sim -14$ \kms \   (blue side) and $\sim 40$ \kms \  (red side).
Note that for CO no high-velocity emission comes from the  positions close to
the protostar.  As for SiO(8--7), the p-v plot shows several features.  Contrary
to the CO case, only very weak SiO emission is coming from the low-velocity
material. The SiO emission resembles the CO emission at distances greater than
$\sim 5''$, with velocities increasing  with distance. However, the most
remarkable characteristic of the SiO p-v plot is that the emission close to the
protostar has the widest  range of velocities, including the highest, up to
$-20$ \kms \ in the blue side and up to 40 \kms \ in the red side. This is a
striking feature of the SiO jet,  that is,  that very high velocities are found
very close ($\sim 500$ AU) to the  protostar. 

A possible explanation for the wide range of velocities found for SiO at the 
spatial scales of the disk would be that SiO close to the protostar is tracing 
a protostellar wind, with a large opening angle, and thus yielding a  maximum
spread of velocities. This interpretation favors a wide-angle wind as the
mechanism for driving the outflow (\eg \ Shu \et  \ 1991), since a pure jet
model, in which velocity vectors point only in the polar direction, cannot
produce this feature (\eg \ Masson \& Chernin 1993; Smith \et \  1997). 
However, the overall structure of the SiO emission is highly collimated, and is
very reminiscent of a pure jet (see Fig.\ \ref{fm0}b,c). In the wind-driven
model, such a collimated structure would be the densest portion of a wide-angle
wind. In this model, at distances $\sim 500$ AU (the position of the  innermost
SiO knots), the density decreases steeply with distance perpendicular to the
axis of the jet, while the velocity vectors still span a wide angle around the
jet axis (see \eg \ Shang \et \ 1998).  Note that the SiO emission close to the
driving source is not due to mixing with entrained material.  The highly
collimated morphology, together with the observed high velocity (up to 40 \kms),
and the derived high density ($> 10^6$ \cmt) and high temperature ($>300$ K,
Hirano \et \ 2005) for the SiO gas very close to the center are strongly
suggestive of material from the primary jet, and not of ambient material being
entrained.

Finally, we also see from the p-v plot that for distances larger than $\sim 5''$
(1500 AU), the velocity of both CO and SiO emission increases with distance.  
This is consistent with high-velocity CO and SiO tracing entrained material,
dragged by the primary jet.  However, the velocities reached by SiO are $\sim 5$
\kms \ higher than those of CO, especially on the red side. This feature is also
seen when superposing the SiO(5--4) p-v plot on the CO(2--1) (Hirano \et \
2005). Presumably, SiO at distances larger than $\sim 5''$ from the protostar
comes from entrained material with higher density than CO. This is consistent
with the critical density of SiO(8--7) being higher than that of CO(3--2). Such
higher-density material would be likely closer to the primary jet, resulting in
entrained SiO showing higher velocities than entrained CO. 

In order to  compare the peak positions of the high-velocity SiO emission with
those of the CO emission,  we  cross correlated the map of high-velocity SiO
with that of high-velocity CO.   The cross correlation function on the blue side
of the jet has a single maximum  at $3\farcs8$, meaning that the SiO knots are
on average $\sim 4''$ closer to the  protostar than the CO knots on the blue
side. For the red side, the cross  correlation function had two maxima, one at
$\sim 1''$, and the other one at  $\sim 6''$, with the SiO knots closer to the
protostar. Thus, for both sides of the jet, the SiO knots are found closer to
the driving source than the CO knots. This suggests that chemical
differentiation can be important in the jet.

In particular, we measured (in the high-velocity maps) the distance to the center
for the SiO knots B1 and R1, and found that the peak is at $\sim 1\farcs5$ (470
AU) from the center. Since B1 and R1 are only slightly resolved by the beam (even
if we include the low-velocities), no significant  emission of SiO is closer to
the protostar than the peak of the innermost knots.

In summary, from the comparison of SiO(8--7) with SiO(5--4) and CO(3--2), it
seems that the SiO(8--7) emission close to the protostar has contributions from
the  primary jet, which could be driven by a wide-angle wind. At projected
distances $\ge 1500$ AU from the protostar, the SiO(8--7) shows velocities
increasing with distance, likely tracing entrained gas. For the same distances,
the CO(3--2) also shows velocities increasing with distance, but reaching
systematically lower velocities than the SiO(8--7). We interpret this feature as
SiO(8--7) tracing entrained gas which is denser, and therefore closer to the
primary jet, than the entrained CO(3--2) gas.

\begin{acknowledgements}
 
We appreciate the collaboration of the SMA staff in the observation and
reduction process, specially from Chunhua Qi and Mark Gurwell.  We wish to
thank Mark McCaughrean and Jenniffer Wiseman for kindly providing the H$_2$ and
\nh \ images, respectively. A. P. is grateful to J. M. Girart for useful
discussions. A. P. and R. E. are supported by a MEC grant AYA2005-08523 and
FEDER funds. H. B. acknowledges financial support by the Emmy-Noether-Program
of the Dutsche Forschungsgemeinschaft (DFG, grant BE2578).

\end{acknowledgements}

\end{document}